\documentclass[11pt,a4paper]{article}

\usepackage[utf8]{inputenc}
\usepackage[T1]{fontenc}
\usepackage{lmodern}

\usepackage{amsmath,amssymb,amsthm}
\usepackage{mathtools}

\mathtoolsset{showonlyrefs=true}
\usepackage{booktabs}
\usepackage{array}
\usepackage{tikz}
\usetikzlibrary{arrows.meta,positioning}
\usepackage{microtype}
\usepackage[numbers,square,sort&compress]{natbib}
\usepackage[font=small,labelfont=bf,labelsep=space]{caption}
\usepackage[section]{placeins}

\usepackage{graphicx}
\usepackage{xcolor}
\usepackage{float}
\usepackage{algorithm}
\usepackage{algorithmic}
\usepackage{comment}

\usepackage[margin=2.65cm,top=2.5cm,bottom=2.65cm]{geometry}
\usepackage{authblk}
\usepackage[colorlinks=true,linkcolor=black,citecolor=black,urlcolor=black]{hyperref}

\newcommand{\R}{\mathbb{R}}
\newcommand{\E}{\mathbb{E}}

\newcommand{\F}{\mathcal{F}}
\newcommand{\A}{\mathcal{A}}

\providecommand{\keywords}[1]{\small\textbf{\textit{Keywords---}} #1}

\def \N{\mathbb{N}}
\def \M{\mathbb{M}}
\def \Z{\mathbb{Z}}
\def \G{\mathbb{G}}
\def \P{\mathbb{P}}
\def \Q{\mathbb{Q}}
\def \W{\mathbb{W}}
\def \d{\mathrm{d}}

\newtheorem{Lemma}{Lemma}[section]

\numberwithin{equation}{section}

\newtheorem{proposition}{Proposition}[section]

\newtheorem{remark}{Remark}[section]
\newtheorem{example}{Example}[section]

\title{Media Trading with Censored Auction Prices:\\ A Stochastic Control and Reinforcement Learning Approach}

\author[1,2]{Yun-Ithry Gamrani}
\author[1]{Huy\^en Pham}

\affil[1]{\'Ecole Polytechnique, Palaiseau, France}
\affil[2]{Siggmund, Paris, France}

\affil[ ]{\texttt{charles-joseph.gamrani@polytechnique.edu}\\
\texttt{huyen.pham@polytechnique.edu}}

\date{}

\begin{document}

\maketitle

\begin{abstract}
In real-time bidding, marketing agencies must balance campaign performance, transaction margins and the use of a finite advertising budget. A key difficulty is that market-clearing prices are typically observed only when bids are successful, so the information available for subsequent decisions depends on the bidding policy. We formulate this problem using a continuous-time stochastic control model with a latent price process. To construct a price signal, we consider parallel exploratory subcampaigns in which bids are randomized subject to an upper bound. Under a stylized cross-sectional sampling model, the resulting price estimator is consistent and satisfies a mean-square error bound of order $N^{-1}$, where $N$ is the number of exploratory observations. We then describe a model-based approach combining forward-equation calibration with Hamilton--Jacobi--Bellman policy iteration and a physics-informed neural network, alongside an entropy-regularized actor--critic approach based on observable information in a model free setting. Synthetic experiments illustrate the estimator and the computational procedures. The framework clarifies the decisions and information trade-offs faced by the agency, while distinguishing full-information benchmarks and estimated-price policies from the genuinely partially observed control problem.
\end{abstract}

\vspace{2mm}





\section{Introduction}
\label{sec:introduction}

Real-time bidding (RTB) has made advertising inventory a resource that must be purchased through frequent, automated auctions. For a marketing agency, the decision is not simply how much to bid on an individual impression: the bids determine campaign performance, the revenue associated with media trading, and the depletion of the client's budget over the campaign horizon. Market prices and campaign outcomes are uncertain, and the agency must adjust its decisions as information arrives. This makes media trading a dynamic resource-allocation problem with an important operational constraint: the agency may not observe the clearing prices of auctions that it loses.

\begin{figure}[htbp]
    \centering
    \includegraphics[width=10cm,height=5.6cm]{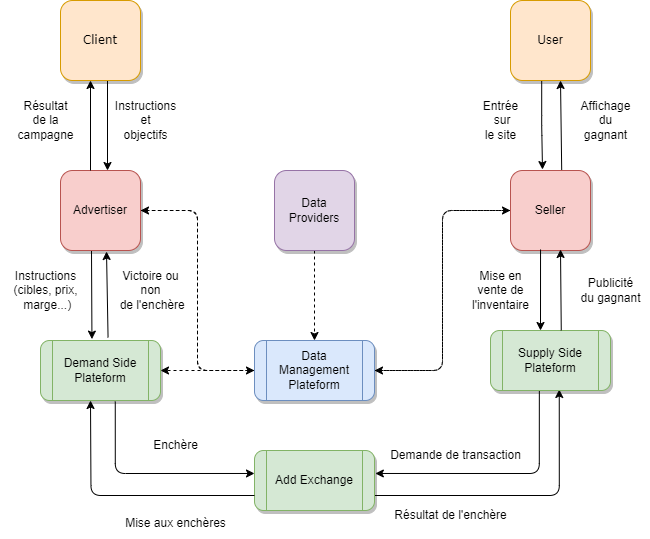}
    \caption{Main participants in a programmatic advertising auction.}
    \label{fig:inter}
\end{figure}

The observation constraint changes the nature of the optimization. A high bid is more likely to generate an impression and reveal the clearing price, but may also consume more of the available budget. A low bid preserves expenditure yet produces less information about the market. Thus, data collection cannot be treated independently of the bidding policy. Figure~\ref{fig:inter} summarizes the information and decision flow.


The problem has two complementary aspects. From a modeling perspective, one needs to represent the market price, campaign performance and remaining budget, together with an observation mechanism that depends on the bid. From a computational perspective, the agency needs implementable policies despite imperfect information and uncertain dynamics. Stochastic control provides a natural description of the intertemporal objectives and constraints; reinforcement learning offers an alternative route to policy optimization when the dynamics are not fully calibrated.

\paragraph{Related work.}
Earlier studies of advertising effectiveness include the analysis of exposure and memory \citep{zielske1959remembering,zielske1980remembering}, aggregate advertising models \citep{little1979aggregate}, and media-mix measurement \citep{52624}. Uncertainty in attributing advertising outcomes remains a major practical difficulty \citep{lewis2015unfavorable,mulc2025nnnnextgenerationneuralnetworks}.

Auction-based advertising has been studied using constrained optimization \citep{Zhang2014OptimalRT}, online learning with side information \citep{Flajolet2017RealTimeBW}, and deep reinforcement learning \citep{Zhao2018DeepRL}. Continuous-time stochastic control approaches to bidding have been proposed by \citep{fernandeztapia2016optimalrealtimebiddingstrategies} and \citep{motte2021optimalbiddingstrategiesdigital}; mean-field models also describe aggregate advertising effects \citep{Ricciardi2024MeanFG}. The statistical problem of estimating auction-clearing prices from censored observations is related to \citep{Wu2015PredictingWP}.

Of particular relevance to the operational setting is the recent contribution in  \citep{Majima2024BayesianBidding}. They study keyword-level bid optimization in sponsored-search advertising using limited aggregate feedback, a Bayesian model of campaign metrics, and a bandit-based online optimization method. Their study includes simulations using real advertising data. The present work addresses a different decision problem: continuous-time bidding for a media-trading agency, explicit censoring of clearing prices by auction outcomes, and a balance between performance, transaction margins and remaining budget. Accordingly, the informational assumptions, decision frequency, and objectives differ; we do not claim a direct empirical comparison with their method.

\paragraph{Modeling and methodological contributions.}
We first set out an agency-level stochastic decision model in which the market price is latent and bidding affects the information revealed by transactions. To construct a usable price signal, we consider randomized bids in parallel exploratory subcampaigns. In an explicit cross-sectional observation model, the winning frequency leads to a consistent estimator with a finite-sample mean-square error bound.

We next connect this price signal to two computational approaches. The first combines forward-equation calibration with a full-information Hamilton--Jacobi--Bellman (HJB) benchmark, approximated through policy iteration and physics-informed neural networks (PINNs) \citep{Raissi2019}. The second describes an entropy-regularized actor--critic method, defined on observable features and inspired by continuous-time reinforcement learning \citep{JMLR:v21:19-144,JMLR:v23:21-0947,jia_q-learning_2022}. These are established numerical ingredients; the interest of combining them here is to clarify their use and limitations for the media-trading decision problem. In particular, the estimated-price feedback is a \emph{plug-in} strategy, not the solution of the underlying partially observed control problem.

The paper is organized as follows. Section~\ref{sec:intro} describes the RTB decision problem and a simple benchmark. Section~\ref{sec:general_model} introduces the stochastic model and the observation structure. Estimation and calibration are considered in Section~\ref{sec:partial_observation_calibration}. Sections~\ref{sec:optimal_control} and~\ref{sec:reinforcement_learning} present the HJB-based and reinforcement-learning approaches. Section~\ref{sec:numerics} discusses the numerical experiments and their limitations, before the concluding section.

\section{The media trading decision problem}
\label{sec:intro}
We begin with the agency's decisions, observations and budget accounting, then use an elementary linear--quadratic example to illustrate how the auction-winning constraint enters the optimization. The example is a full-information benchmark; the incomplete-information model follows in the next section.

\subsection{Auction mechanism and information}

Consider an advertising campaign over a finite horizon $[0,T]$. The agency receives an initial budget from its client and submits bids to a sequence of RTB auctions through a demand-side platform.

At time $t$, the agency chooses a bid $\alpha_t$. An auction is won if $\alpha_t\ge S_t$, where $S_t$ denotes the market-clearing price. A successful bid contributes to campaign performance and consumes part of the budget.

The agency observes the clearing price only when it wins. Its observation history thus depends on the previous bids, and the decision problem combines partial observation with control-dependent information acquisition.

Following the continuous-time approaches of \citep{fernandeztapia2016optimalrealtimebiddingstrategies} and \citep{motte2021optimalbiddingstrategiesdigital}, we use a diffusion approximation to represent the frequent auctions and the resulting campaign dynamics.

We use the following variables:
\begin{itemize}
    \item $S_t$, the latent market-clearing price;
    \item $X_t$, the campaign performance (for instance the cumulative number of conversions or sales);
    \item $Y_t$, the remaining campaign budget;
    \item $\alpha_t$, the bid, which is the control rather than a state variable.
\end{itemize}

The trade-offs have a direct operational interpretation. Raising the bid increases the chance of obtaining an impression, while successful transactions affect both campaign value and the agency's margin. The remaining budget couples today's bid with tomorrow's opportunities. Moreover, a lost auction reveals less price information than a won auction. We first examine these mechanisms in a simple complete-information model; a joint optimization of exploration expenditure is beyond the scope of the present framework.

\subsection{A linear--quadratic benchmark}

As an elementary illustration, we first consider a first-price stylization in which a winning bid is paid at its submitted value. Unlike the market-clearing-price settlement in the general model below, this convention leads to a particularly transparent quadratic maximization. Suppose that the market price is fully observed and follows the Ornstein--Uhlenbeck dynamics
\begin{align} \label{eq:dynamics}   
dS_t &= \mu(\bar S-S_t)\,dt+\sigma\,dW_t^{1}. 
\end{align} 
The agency controls its bidding price $\alpha_t\in A=[0,A_{\max}]$. An auction is won whenever
\[
\alpha_t\ge S_t.
\]
Campaign performance evolves according to
\[
\left\{
\begin{aligned}
dX_t
    &=\gamma\,
      \mathbf 1_{\{\alpha_t\ge S_t\}}dt
      +\theta\,dW_t^{2},\\
X_0&=0,
\end{aligned}
\right.
\]
where $\gamma$ measures the average increase in performance generated by a successful auction.
The remaining budget satisfies
\[
\left\{
\begin{aligned}
dY_t
    &=-(1+\beta)\alpha_t
      \mathbf1_{\{\alpha_t\ge S_t\}}dt,\\
Y_0&=B,
\end{aligned}
\right.
\]
where $\beta>0$ denotes the agency margin.

The agency seeks to maximize the performance criterion

\[
J(\alpha)
=
\mathbb E
\left[
\int_{0}^{T}
\left(
X_t
+
\beta\alpha_t
\mathbf1_{\{\alpha_t\ge S_t\}}
-
\kappa(1+\beta)^2
\alpha_t^2
\mathbf1_{\{\alpha_t\ge S_t\}}
\right)
dt
-
Y_T^2
\right].
\]

The running reward combines campaign performance, the margin obtained from successful bids, and a quadratic penalty on the bidding cost. The terminal term penalizes any deviation of the final budget from zero, including an overdraft. This is a soft penalty, not a hard budget constraint.

Let $p=(p_s,p_x,p_y)\in\mathbb R^3$ and $M=(M_{ij})_{i,j\in\{s,x,y\}}\in\mathbb S^3$ be generic variables. The (maximized) Hamiltonian of this benchmark is defined by
\begin{equation}\label{eq:hamiltonian_LQ}
\begin{aligned}
H_{\rm LQ}(s,x,y,p,M)
={}&\mu(\bar S-s)p_s
+\frac{\sigma^2}{2}M_{ss}
+\frac{\theta^2}{2}M_{xx}+x\\
&+\sup_{a\in[0,A_{\max}]}
\mathbf1_{\{a\ge s\}}
\Big[\gamma p_x
+\bigl(\beta-(1+\beta)p_y\bigr)a
-\kappa(1+\beta)^2a^2\Big].
\end{aligned}
\end{equation}
Here $\mathbb S^3$ is the space of symmetric $3\times3$ matrices. The state variable $y$ does not appear explicitly in the present coefficients, but is retained as an argument of the Hamiltonian. For a smooth value function $V$, the arguments $(p,M)$ in \eqref{eq:hamiltonian_LQ} are replaced by $(DV,D^2V)$, where $DV=(V_s,V_x,V_y)$. The HJB equation therefore reads
\begin{equation}\label{eq:HJB_LQ}
\left\{
\begin{aligned}
&V_t+H_{\rm LQ}(s,x,y,DV,D^2V)=0,\\
&V(T,s,x,y)=-y^2.
\end{aligned}
\right.
\end{equation}

The optimization over the bid can be performed explicitly, taking the winning constraint into account. For generic $p\in\mathbb R^3$, set
\[
\begin{aligned}
Q_p(a)&=\gamma p_x+\bigl(\beta-(1+\beta)p_y\bigr)a
-\kappa(1+\beta)^2a^2,\\
a_0(p)&=\frac{\beta-(1+\beta)p_y}{2\kappa(1+\beta)^2}.
\end{aligned}
\]
If $s\le A_{\max}$, the best bid conditional on winning is
\[
a_{\rm win}(s,p)
=\operatorname{proj}_{[\max\{0,s\},A_{\max}]}a_0(p).
\]
Writing $B(s,x,p,M)=\mu(\bar S-s)p_s+\frac{\sigma^2}{2}M_{ss}+\frac{\theta^2}{2}M_{xx}+x$ and $a_{\rm win}=a_{\rm win}(s,p)$, we obtain
\begin{equation}\label{eq:hamiltonian_LQ_explicit}
H_{\rm LQ}(s,x,y,p,M)=B(s,x,p,M)+
\begin{cases}
0, & s>A_{\max},\\
Q_p(a_{\rm win}), & s\le0,\\
\max\{0,Q_p(a_{\rm win})\}, & 0<s\le A_{\max}.
\end{cases}
\end{equation}
For $0<s\le A_{\max}$, the optimal bid wins the auction only if $Q_p(a_{\rm win})\ge0$; otherwise any bid below $s$ avoids the transaction. If $s>A_{\max}$, no admissible bid wins the auction. In these formulas the generic vector $p$ is evaluated at $DV(t,s,x,y)$ when defining the optimal feedback; the matrix $M$ is evaluated at $D^2V$ in the HJB equation.

\begin{figure}[htbp]
    \centering
    \includegraphics[width=0.90\linewidth]{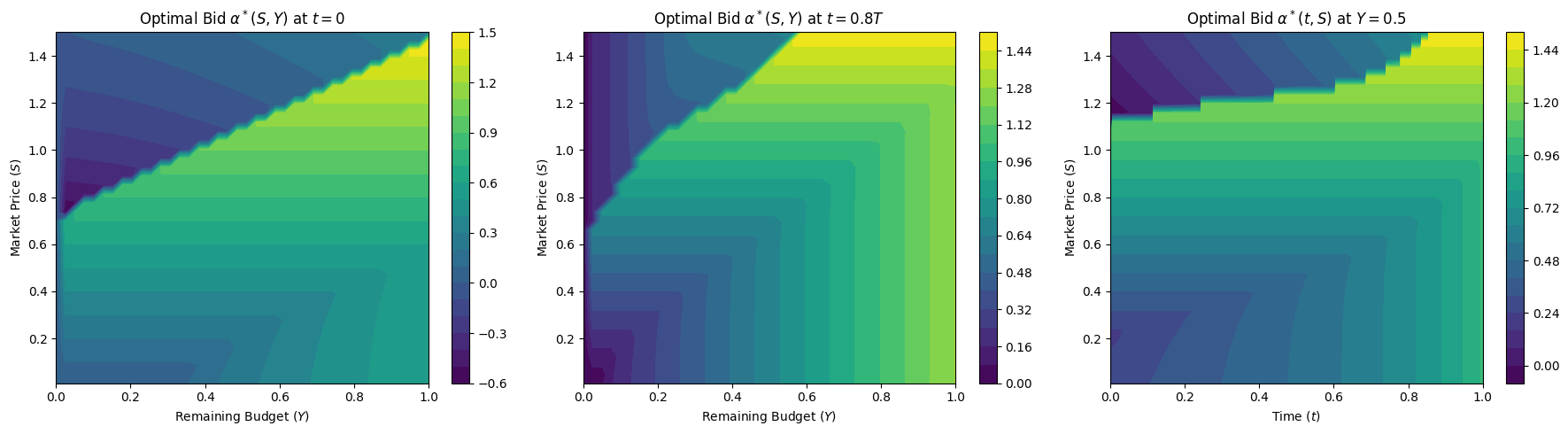}
    \caption{Bidding policy from the benchmark implementation}
    \label{fig:alphaopt}
\end{figure}

\begin{figure}[htbp]
    \centering
    \includegraphics[width=0.90\linewidth]{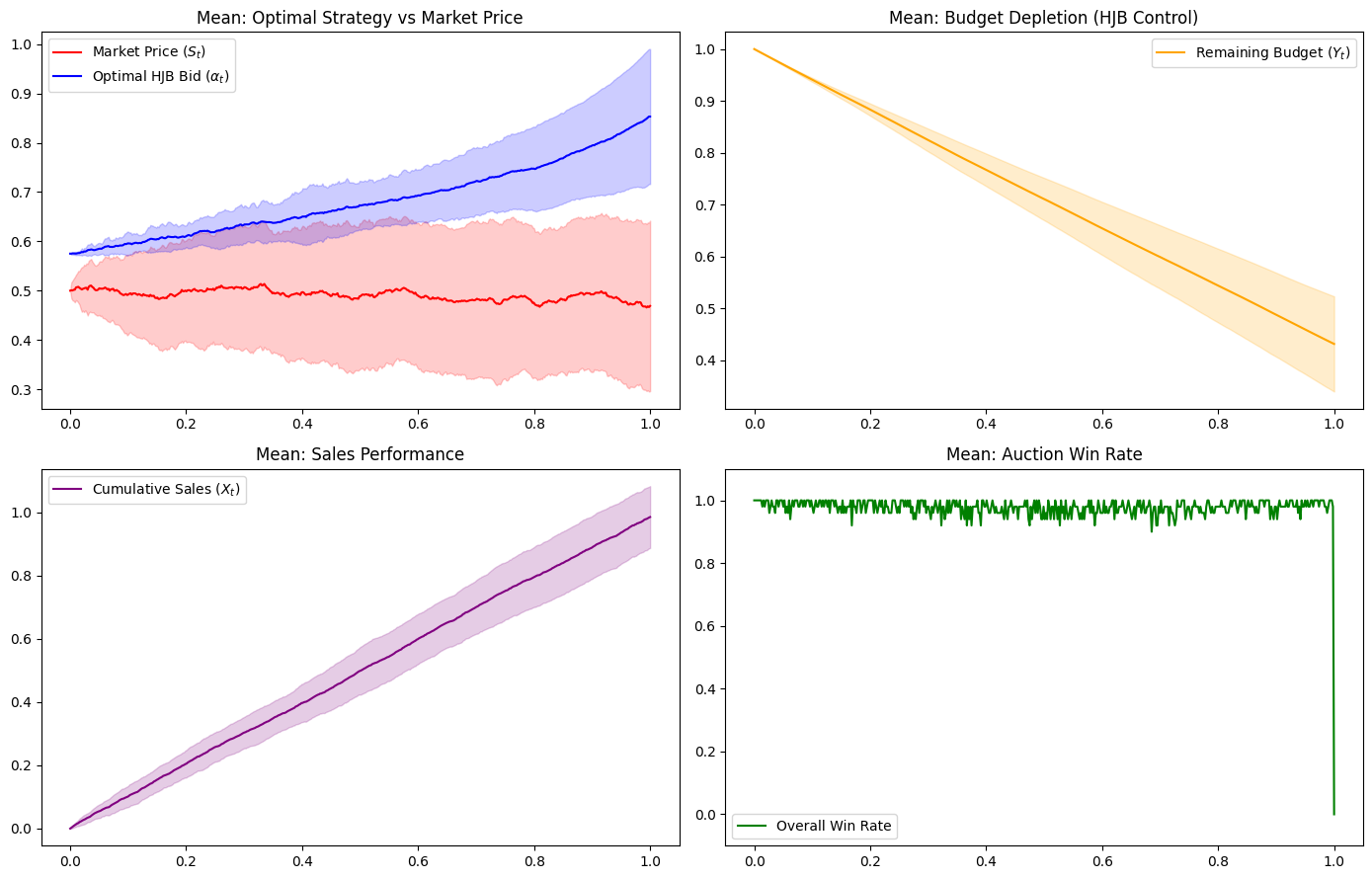}
    \caption{Benchmark trajectories}
    \label{fig:trajoptlq}
\end{figure}

Figures~\ref{fig:alphaopt} and~\ref{fig:trajoptlq} illustrate the bidding and budget trajectories produced by the numerical implementation.

The preceding calculations rely on observing $(S_t)$ continuously. We now turn to the actual information structure, in which the market price is hidden except for the transactions won by the agency.

\section{Stochastic dynamics and information structure}
\label{sec:general_model}

We now introduce the general stochastic control framework studied in the remainder of the paper. Contrary to the benchmark model of Section~\ref{sec:intro}, we allow for general nonlinear dynamics of the market-clearing price and of the campaign performance. The presentation proceeds in three steps: we first formulate the problem under \emph{full information}, that is, as if the agency had access to the whole filtration $\mathbb{F}$ (Section~\ref{subsec:full_information}); we then explain how the agency can enrich this information by running auxiliary sub-campaigns, and we state the problem addressed in this paper: constructing an approximate solution of the control problem from these limited observations (Section~\ref{subsec:subcampaigns}).

\subsection{The full-information model}
\label{subsec:full_information}

Let $(\Omega,\mathcal F,\mathbb{F}=(\mathcal F_t)_{t\in[0,T]},\mathbb P)$ be a filtered probability space satisfying the usual conditions and supporting two independent one-dimensional $\F$-Brownian motions $W^1$ and $W^2$. The filtration $\F$ represents the full information of the model: it contains every source of randomness affecting the market and the campaign, including quantities that the agency does not observe. It serves as a reference framework, in which all processes below are defined; the information actually available to the agency is described in Section~\ref{subsec:randomized_observation}.

The latent market-clearing price $S=(S_t)_{t\in[0,T]}$ is the solution of
\begin{equation}
\label{eq:S_general}
dS_t=\mu(t,S_t)\,dt+\sigma(t,S_t)\,dW^1_t,
\qquad S_0=s_0,
\end{equation}
where $s_0$ is an $\mathcal F_0$-measurable, square-integrable random variable. The functions $\mu$ and $\sigma$ describe the local trend and the volatility of the market-clearing price. The price is exogenous: the agency's bids are assumed to be too small, relative to the size of the market, to affect $S$.

The agency controls its bidding strategy through a process $\alpha=(\alpha_t)_{t\in[0,T]}$ taking values in the compact set
\[
A=[0,A_{\max}],\qquad A_{\max}>0.
\]
In the full-information framework, the set of bidding strategies is
\[
\mathcal A^{\mathbb{F}}
=
\Big\{\alpha=(\alpha_t)_{t\in[0,T]}\ :\ \alpha \text{ is } \mathbb{F}\text{-progressively measurable and } A\text{-valued}\Big\}.
\]
Since $A$ is bounded, every $\alpha\in\mathcal A^{\mathbb{F}}$ satisfies $\mathbb E\big[\int_0^T\alpha_t^2\,dt\big]<\infty$, so that no additional integrability condition is required. All processes introduced below depend on the chosen strategy $\alpha$, which we indicate by a superscript.

At time $t$, the auction is won if and only if the submitted bid exceeds the market-clearing price, that is, on the event $\{\alpha_t\ge S_t\}$. The cumulative number of auctions won is thus given by
\begin{equation}
\label{eq:Z_general}
Z^\alpha_t=\int_0^t\mathbf 1_{\{\alpha_u\ge S_u\}}\,du,
\qquad t\in[0,T].
\end{equation}

The performance process $X^\alpha=(X^\alpha_t)_{t\in[0,T]}$ of the campaign represents, depending on the application, the cumulative number of clicks, conversions or sales generated by the campaign. It evolves according to
\begin{equation}
\label{eq:X_general}
dX^\alpha_t
=
\gamma(t,X^\alpha_t,S_t,\alpha_t)\,dt
+
\theta(t,X^\alpha_t,S_t,\alpha_t)\,dW^2_t,
\qquad X^\alpha_0=x_0\in\mathbb R.
\end{equation}
The drift and diffusion coefficients may depend simultaneously on the current market conditions and on the bid selected by the agency.

The agency starts the campaign with an initial budget $B_0>0$, and $Y^\alpha=(Y^\alpha_t)_{t\in[0,T]}$ denotes the remaining budget. Whenever an auction is won, the agency pays the market-clearing price together with its trading margin. For a fixed margin parameter $\beta>0$, the budget dynamics are
\begin{equation}
\label{eq:Y_general}
Y^\alpha_t
=
B_0-(1+\beta)\int_0^t S_u\,\mathbf 1_{\{\alpha_u\ge S_u\}}\,du,
\qquad t\in[0,T].
\end{equation}

Throughout the following, we assume that:
\begin{enumerate}
    \item[(H1)] $\mu,\sigma:[0,T]\times\mathbb R\to\mathbb R$ are Borel measurable, Lipschitz continuous in $s$ uniformly in $t$, and of linear growth in $s$;
    \item[(H2)] $\gamma,\theta:[0,T]\times\mathbb R\times\mathbb R\times A\to\mathbb R$ are Borel measurable, Lipschitz continuous in $x$ uniformly in $(t,s,a)$, and satisfy
    \[
    |\gamma(t,x,s,a)|+|\theta(t,x,s,a)|\le C\,(1+|x|+|s|)
    \]
    for some constant $C>0$;
    \item[(H3)] $f:A\times\mathbb R\times\mathbb R\to\mathbb R$ and $g:\mathbb R\to\mathbb R$ are Borel measurable and of quadratic growth:
    $|f(a,s,x)|\le C(1+|s|^2+|x|^2)$ and $|g(y)|\le C(1+|y|^2)$.
\end{enumerate}
Under (H1)--(H2), equation~\eqref{eq:S_general} admits a unique strong solution and, for every $\alpha\in\mathcal A^{\mathbb{F}}$, equation~\eqref{eq:X_general} (whose coefficients are random through $S$ and $\alpha$) admits a unique strong solution $X^\alpha$, with
$\mathbb E\big[\sup_{t\le T}(|S_t|^2+|X^\alpha_t|^2)\big]<\infty$; see, e.g.,~\cite{pham2009continuous}. The processes $Z^\alpha$ and $Y^\alpha$ are then well defined, continuous and $\mathbb{F}$-adapted, and $\mathbb E[|Y^\alpha_T|^2]<\infty$.

The agency aims at jointly maximizing campaign performance and trading revenue, while ensuring that the available budget is efficiently spent over the campaign horizon. For $\alpha\in\mathcal A^{\mathbb{F}}$, we define the performance criterion
\begin{equation}
\label{eq:criterion}
J(\alpha)
=
\mathbb E\left[
\int_0^T
\Big(
\beta S_t\,\mathbf 1_{\{\alpha_t\ge S_t\}}
+
f(\alpha_t,S_t,X^\alpha_t)
\Big)\,dt
-
g(Y^\alpha_T)
\right],
\end{equation}
which is well defined under (H1)--(H3). The function $f$ measures the instantaneous performance of the campaign: for instance, $f(a,s,x)=x$ corresponds to maximizing the cumulative number of sales, while other choices model cost-per-click or cost-per-acquisition objectives. The terminal function $g$ penalizes unspent budget, e.g.\ $g(y)=y^2$, and reflects the practical requirement of spending the campaign budget before the end of the planning horizon. The \emph{full-information problem} is
\begin{equation}
\label{eq:control_problem}
V^{\mathbb{F}}=\sup_{\alpha\in\mathcal A^{\mathbb{F}}}J(\alpha).
\end{equation}
Problem~\eqref{eq:control_problem} is a standard Markovian stochastic control problem in the state variables $(S,X,Y)$. It is, however, not the problem faced by the agency: an optimal strategy for~\eqref{eq:control_problem} generally requires the knowledge of the current price $S_t$, which is not available in practice. The value $V^{\mathbb{F}}$ will therefore only be used as a benchmark.
\subsection{Sub-campaigns and approximate problem}
\label{subsec:subcampaigns}

To overcome this lack of information, the agency may allocate part of its budget to $N$ auxiliary sub-campaigns. This possibility is at the core of the randomization procedure developed in this paper. The purpose of the sub-campaigns is not to optimize their own results, but to generate observations of the market price that will be used to control the main campaign. They are described in more detail in Section~\ref{sec:partial_observation_calibration}.

Each sub-campaign $i\in\{1,\dots,N\}$ faces its own market-clearing price $S^i=(S^i_t)_{t\in[0,T]}$, a continuous, $\mathbb{F}$-progressively measurable process obtained as a perturbed version of $S$; its construction is detailed in Section~\ref{subsec:randomized_observation}. 
On sub-campaign $i$, the agency uses a prescribed $A$-valued
randomized bidding process
$\alpha^i=(\alpha_t^i)_{t\in[0,T]}$, which is not subject to
optimization. In the exploration scheme of Section~\ref{subsec:randomized_observation}, the bids
are generated independently of the latent price process $S$
and of the subcampaign price noises. The reference filtration
$\mathbb F$ is understood to include this exogenous randomization.
The cumulative winning-time process on sub-campaign $i$ is
\begin{equation}
\label{eq:Z_alpha}
Z^{i}_t=\int_0^t\mathbf 1_{\{\alpha^i_u\ge S^i_u\}}\,du,
\qquad t\in[0,T].
\end{equation}

 Our objective is then to construct an approximate solution of \eqref{eq:control_problem} from the limited information available, that are $X^\alpha$ and $(Z^i)_{i \in \mathbb{N}}$. To this end, we use the sub-campaign observations to build a proxy $\widehat S$ of the latent price $S$ (Section~\ref{subsec:randomized_observation}). Substituting $\widehat S$ for $S$ in the solution of the full-information problem~\eqref{eq:control_problem} yields a strategy in $\mathcal A$, whose performance can be compared with the benchmark value $V^{\mathbb{F}}$.
\section{Price estimation and model calibration}
\label{sec:partial_observation_calibration}

The operational model raises two distinct estimation questions. The current market price must be inferred from censored auction outcomes, while the dynamics of prices and campaign performance have to be fitted to historical data. We address these questions separately, starting with the information obtained from randomized exploratory bids. This separation makes the resulting method tractable but does not optimize exploration and exploitation jointly.

\subsection{Exploration through randomized bids}
\label{subsec:randomized_observation}

Consider $N$ exploratory subcampaigns with a common latent market factor $S$. At each observation time, the price of subcampaign $i$ is
\[
S_t^i=S_t+\varepsilon_t^i,
\qquad \varepsilon_t^i\sim N(0,v_t^2),
\]
where $v_t>0$ is known and the noises are conditionally independent across subcampaigns. This is a stylised cross-sectional sampling model, not an assertion that auction prices are exactly Gaussian in practice.

The main-campaign bids belong to $A=[0,A_{\max}]$. To respect the same bidding cap during exploration, we draw exploratory bids independently from an exponential distribution \emph{conditioned on} $[0,A_{\max}]$:
\begin{equation}\label{eq:exploration_density}
q_\lambda(a)=\frac{\lambda e^{-\lambda a}}{1-e^{-\lambda A_{\max}}}
\mathbf 1_{[0,A_{\max}]}(a),\qquad \lambda>0.
\end{equation}
The exploratory bids are independent of $S$ and of the price noises. The agency observes the winning indicators, hence the empirical winning fraction
\begin{equation}\label{eq:ZN_definition}
z_t^N=\frac1N\sum_{i=1}^N\mathbf 1_{\{\alpha^i\ge S_t+\varepsilon_t^i\}}.
\end{equation}
Writing $\Phi$ for the standard Gaussian distribution function, the conditional winning probability is
\begin{equation}\label{eq:observation_map}
h_t(s):=\mathbb P(\alpha^i\ge s+\varepsilon_t^i\mid S_t=s)
=\int_0^{A_{\max}}q_\lambda(a)
\Phi\!\left(\frac{a-s}{v_t}\right)\,da.
\end{equation}
In particular, $h_t$ is a continuous strictly decreasing function from $\mathbb R$ onto $(0,1)$, since
\begin{equation}\label{eq:observation_derivative}
h_t'(s)=-\frac1{v_t}\int_0^{A_{\max}}q_\lambda(a)
\phi\!\left(\frac{a-s}{v_t}\right)\,da<0,
\end{equation}
where $\phi$ is the standard Gaussian density. The empirical win rate can therefore be inverted to estimate the hidden price. The assumption of an exogenous independent exploration channel is important: it allows elementary statistical estimation, but does not solve the joint exploration--exploitation problem for the full campaign.

\begin{remark}
The mean-square error bound in Lemma~4.1 assumes that the latent
price lies almost surely in a fixed compact interval. For the
Ornstein--Uhlenbeck model considered in Section~7, this assumption
does not hold, and the projection in~\eqref{eq:price-estimator} introduces a truncation
error that is not covered by the bound.
\end{remark}

\subsection{Estimation error and stability}
\label{subsec:estimator_convergence}

\begin{proposition}\label{prop:observation_convergence}
For each fixed $t$ with $v_t>0$, suppose that the pairs
$(\alpha^i,\varepsilon_t^i)$ are conditionally i.i.d. given $S_t$,
with the distributions specified above. Then
\[
\mathbb E[z_t^N\mid S_t]=h_t(S_t),\qquad
\mathbb E[|z_t^N-h_t(S_t)|^2]\le\frac1{4N},
\]
and $z_t^N\to h_t(S_t)$ in $L^2$ and probability. If the noise scales $v_t$ are measurable and strictly positive, then
\[
\mathbb E\int_0^T|z_t^N-h_t(S_t)|^2\,dt\le\frac{T}{4N}.
\]
In particular, these assertions concern fixed-time and integrated mean-square convergence.
\end{proposition}

Let $K=[s_-,s_+]$ be a compact interval containing the latent price. We define the inverse estimator
\begin{equation}\label{eq:price-estimator}
\widehat S_t^N=
\left(h_t\big|_K\right)^{-1}\!\left(
\operatorname{proj}_{[h_t(s_+),h_t(s_-)]}z_t^N\right).
\end{equation}
The projection in \eqref{eq:price-estimator} makes the estimator well-defined even when the empirical winning fraction is zero or one. Since $h_t$ is strictly decreasing for $v_t>0$, the inverse is unique; compactness provides a uniform stability bound.

\begin{Lemma}\label{lemma1}
Suppose $S_t\in K$ almost surely, and
\[
m_t:=\inf_{s\in K}|h_t'(s)|>0.
\]
Then
\[
\mathbb E|S_t-\widehat S_t^N|^2
\le \frac{1}{4m_t^2N}.
\]
If $\inf_{0\le t\le T}m_t=m>0$, then
$\mathbb E\int_0^T|S_t-\widehat S_t^N|^2dt\le T/(4m^2N)$.
\end{Lemma}
\begin{proof}
Conditional on $S_t$, $Nz_t^N$ is binomial with success probability
$h_t(S_t)$, and therefore its conditional variance is at most
$N/4$. Projection onto an interval does not increase distance to a
point in that interval. On $K$ the inverse of $h_t$ is
$m_t^{-1}$-Lipschitz. The result follows by conditioning on $S_t$.
\end{proof}

\begin{proposition}\label{prop:payoff_approximation}
Fix a strategy $\alpha$ that is admissible for the original
observation problem. Let $I_t=\mathbf1_{\{\alpha_t\ge S_t\}}$ and
$\widehat I_t=\mathbf1_{\{\alpha_t\ge\widehat S_t^N\}}$, and define
\[
Y_T=B_0-(1+\beta)\int_0^T S_tI_t\,dt,
\qquad
\widehat Y_T^N=B_0-(1+\beta)\int_0^T
\widehat S_t^N\widehat I_t\,dt.
\]
Let $J(\alpha)$ denote the payoff in \eqref{eq:criterion},
and let $J^N(\alpha)$ be the corresponding auxiliary payoff with $S$, $I$ and $Y_T$
replaced by $\widehat S^N$, $\widehat I$ and $\widehat Y_T^N$
(the performance process $X$ is held fixed in this comparison).
Suppose the estimates in Lemma~\ref{lemma1} hold uniformly in time,
$S_t,\widehat S_t^N$ take values in a compact interval $K$, and
$f$ and $g$ are Lipschitz on the relevant ranges. Assume, moreover,
that for all $\delta>0$ small enough
\begin{equation}\label{eq:anti-concentration}
\mathbb P(|S_t-\alpha_t|\le\delta)\le M_\alpha\delta
\qquad\text{for a.e. }t\in[0,T],
\end{equation}
with $M_\alpha$ independent of $N$. Then
\[
|J(\alpha)-J^N(\alpha)|\le C_\alpha N^{-1/3},
\]
where $C_\alpha$ is independent of $N$.
\end{proposition}
\begin{proof}
For each $\delta>0$,
\[
|I_t-\widehat I_t|
\le\mathbf1_{\{|S_t-\alpha_t|\le\delta\}}
+\mathbf1_{\{|S_t-\widehat S_t^N|>\delta\}}.
\]
Markov's inequality, \eqref{eq:anti-concentration} and the
mean-square estimator bound imply
\[
\mathbb E|I_t-\widehat I_t|
\le M_\alpha\delta+\frac{C}{N\delta^2}.
\]
Take $\delta=N^{-1/3}$. Since both price variables remain bounded,
\[
\mathbb E|S_t I_t-\widehat S_t^N\widehat I_t|
\le \mathbb E|S_t-\widehat S_t^N|+
\sup_{s\in K}|s|\,\mathbb E|I_t-\widehat I_t|.
\]
Integrating this inequality and applying the Lipschitz assumptions
to $f$ and $g$ yields the asserted bound.
\end{proof}

\paragraph{Implications for optimal control.}
Proposition~\ref{prop:payoff_approximation} concerns a fixed policy satisfying the anti-concentration assumption. It does not provide a uniform estimate over arbitrary observation-adapted policies. Indeed, the choice $\alpha_t=\widehat S_t^N$ wins by construction in the proxy model, while its outcome remains uncertain in the true model. Therefore convergence of the price estimator alone does not establish near-optimality of an optimizer of the proxy problem. Such a conclusion requires additional stability conditions on the admissible policy class.

In what follows, $\widehat S^N$ is used as an observable price signal in a plug-in control approximation. This is not an exact separation of the partially observed problem. 

\begin{example}[A synthetic check of the price estimator]
We illustrate the estimator of \eqref{eq:price-estimator} in a cross-sectional experiment. Fix $S_t=0.5$, $A_{\max}=1$, $\lambda=3$, $v_t=0.15$, and $K=[0,1]$. For each value of $N$, we generated $3000$ independent samples of exploratory bids and Gaussian price noises, using the capped exponential law \eqref{eq:exploration_density}. Figure~\ref{fig:convergence-Sn} displays the empirical mean-square estimation error. The observed decrease is close to $N^{-1}$, as expected from Lemma~\ref{lemma1}. This numerical check concerns price estimation only, not the performance of a bidding strategy.
\begin{figure}[htbp]
    \centering
    \includegraphics[width=0.78\linewidth]{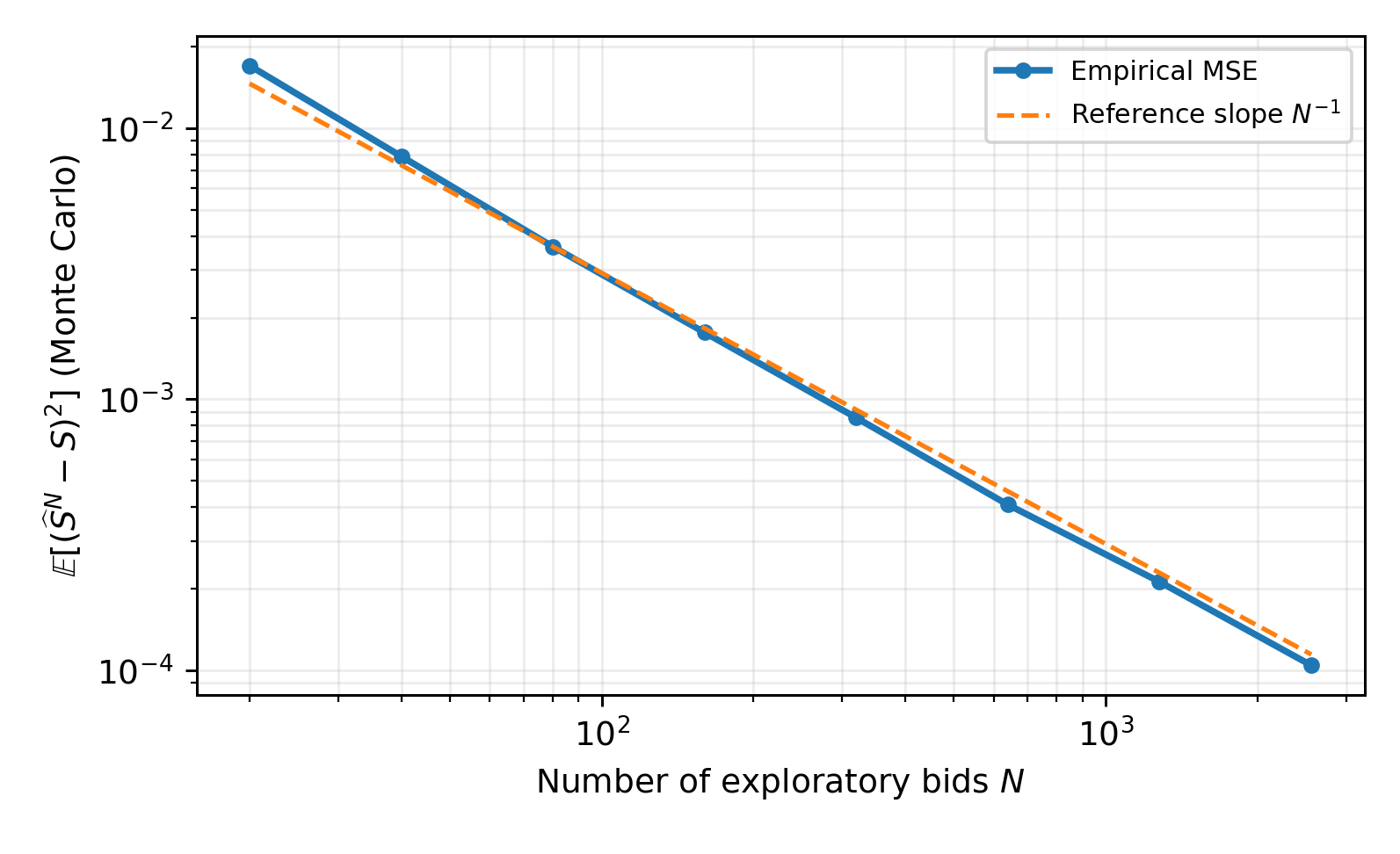}
    \caption{Mean-square error of the capped-bid estimator versus sample size, with $3000$ independent replications and $S=0.5$; dashed reference line of slope $-1$}
    \label{fig:convergence-Sn}
\end{figure}
\end{example}

\subsection{Forward-equation calibration}
\label{subsec:data_driven_calibration}

We now turn to the coefficients that determine the evolution of the price and campaign-performance processes. Even when contemporaneous prices are censored, historical campaign batches may provide data from which to estimate
\[
\mu,\quad \sigma,\quad \gamma,\quad \theta.
\]
As bids affect campaign performance, calibration must take the joint distribution of states and actions into account. The observation or experimental policy must be recorded; otherwise parameters of the performance equation may not be identifiable.

Let
\[
p(t,s,x,a)
\]
denote the joint density of $(S_t,X_t,\alpha_t)$, provided it exists. For deterministic feedback controls, the conditional law of the action may be singular; the corresponding expressions should then be understood in terms of state--action occupation measures. For any test function
$\Delta\in C^{1,2}((0,T)\times\mathbb R^2)$, Itô's formula gives
\[
\begin{aligned}
\mathbb E[\Delta(t,S_t,X_t)]
=&\,
\mathbb E[\Delta(0,S_0,X_0)]
\\
&+
\mathbb E
\left[
\int_0^t
\left(
\partial_r\Delta
+
\mu(r,S_r)\partial_s\Delta
+
\gamma(r,X_r,S_r,\alpha_r)\partial_x\Delta
\right.\right.
\\
&\hspace{3.5cm}
\left.\left.
+
\frac12\sigma^2(r,S_r)\partial_{ss}\Delta
+
\frac12\theta^2(r,X_r,S_r,\alpha_r)\partial_{xx}\Delta
\right)
dr
\right].
\end{aligned}
\]
Writing $\rho(t,s,x)=\int_A p(t,s,x,a)\,da$ for the state marginal, the weak forward equation takes the form
\begin{equation}
\label{eq:controlled_forward_weak}
\begin{aligned}
0
=
\int_A
\Big[
\partial_t p
&+
\partial_s(\mu p)
+
\partial_x(\gamma p)
\\
&-
\frac12\partial_{ss}(\sigma^2p)
-
\frac12\partial_{xx}(\theta^2p)
\Big](t,s,x,a)\,da.
\end{aligned}
\end{equation}

Suppose that historical data provide observations
\[
\mathcal D
=
\left\{
(t_i,X_{t_i}^j,\widehat S_{t_i}^j,\alpha_{t_i}^j):
1\le i\le n,\ 1\le j\le M_i
\right\},
\]
where $\widehat S$ is reconstructed from the collected auction data. We estimate the marginal densities of $S_t$ and $X_t$ using kernel estimators: 
\begin{equation}
\label{eq:kde_price}
\widehat\varphi_i(s)
=
\frac{1}{M_i h_i^S}
\sum_{j=1}^{M_i}
K\left(
\frac{s-\widehat S_{t_i}^j}{h_i^S}
\right),
\end{equation}
and
\begin{equation}
\label{eq:kde_performance}
\widehat\psi_i(x)
=
\frac{1}{M_i h_i^X}
\sum_{j=1}^{M_i}
K\left(
\frac{x-X_{t_i}^j}{h_i^X}
\right).
\end{equation}
Here $K$ is a symmetric kernel satisfying
\[
K\ge0,
\qquad
K(u)=K(-u),
\qquad
\int_{\mathbb R}K(u)\,du=1,
\]
and the bandwidths $h_i^S$ and $h_i^X$ may be selected by cross-validation.

We consider parametric families
\[
\mu^\eta,\quad \sigma^\eta,\quad \gamma^\eta,\quad \theta^\eta,
\]
where $\eta$ is an unknown parameter vector. Calibration amounts to controlling the residual of \eqref{eq:controlled_forward_weak}, together with the discrepancy between observed and model-implied data. We use physics-informed neural networks to approximate the forward densities and their derivatives.

Let
\[
(t_k,s_k,x_k)_{k=1}^P
\]
be collocation points sampled in the state-time domain. The price-density residual is
\begin{equation}
\label{eq:loss_phi}
\mathcal L_\varphi(\eta)
=
\frac1P
\sum_{k=1}^P
\left|
\partial_t\widehat\varphi(t_k,s_k)
+
\partial_s
\left(
\mu^\eta(t_k,s_k)\widehat\varphi(t_k,s_k)
\right)
-
\frac12
\partial_{ss}
\left(
(\sigma^\eta)^2(t_k,s_k)\widehat\varphi(t_k,s_k)
\right)
\right|^2.
\end{equation}

Similarly, the $X$-marginal density
$\psi(t,x)=\int_{\mathbb R}\int_A p(t,s,x,a)\,da\,ds$
satisfies, assuming vanishing boundary fluxes,
\[
\begin{aligned}
0={}&\partial_t\psi(t,x)
+\partial_x\int_{\mathbb R}\!\int_A
\gamma(t,x,s,a)p(t,s,x,a)\,da\,ds\\
&-\frac12\partial_{xx}\int_{\mathbb R}\!\int_A
\theta^2(t,x,s,a)p(t,s,x,a)\,da\,ds.
\end{aligned}
\]
A residual associated with the $X$-marginal equation must integrate over both price and action:
\begin{equation}\label{eq:loss_psi}
\begin{split}
\mathcal L_\psi(\eta)=\frac1P\sum_{k=1}^P\Bigg|
&\partial_t\widehat\psi(t_k,x_k)\\
&+\partial_x\int_{\mathbb R}\int_A
 \gamma^\eta(t_k,x_k,s,a)\widehat p(t_k,s,x_k,a)\,da\,ds\\
&-\frac12\partial_{xx}\int_{\mathbb R}\int_A
 (\theta^\eta)^2(t_k,x_k,s,a)\widehat p(t_k,s,x_k,a)\,da\,ds
\Bigg|^2.
\end{split}
\end{equation}
Here $\widehat p$ estimates the joint state--action occupation density, or an appropriate discretisation of the occupation measure. The time derivatives in the residual require a sufficiently regular interpolation of the empirical densities.

The calibration loss is then
\begin{equation}
\label{eq:calibration_loss}
\mathcal L(\eta)
=
\mathcal L_\varphi(\eta)
+
\mathcal L_\psi(\eta)
+
\mathcal L_{\rm data}(\eta),
\end{equation}
Here $\mathcal L_{\rm data}$ measures agreement with observed campaign data.

A small residual of the forward equation does not, by itself, identify the unknown coefficients: different models may yield the same observed marginal distributions. For completeness, the following elementary local stability bound isolates the additional identifiability assumption; its proof is deferred to Appendix~\ref{app:loss-consistency}.
\begin{proposition}\label{prop:calibration_consistency}
Let $\Theta\subset\mathbb R^d$ be finite-dimensional, and let
$F(\eta)=p^\eta$ denote the model-implied state density as an
element of $H=L^2((0,T)\times\Omega)$. Suppose $F$ is continuously
Fr\'echet differentiable near the true parameter $\eta^\star$ and
\[
\|DF(\eta^\star)h\|_H\ge c_0|h|
\qquad\text{for all }h\in\mathbb R^d,
\]
for some $c_0>0$. Suppose also that the forward equations admit
a uniformly bounded evolution family on $L^2(\Omega)$.
For an approximate density $\widehat p$ with initial data
$\widehat p_0$, define the forward residual
$R^\eta=\partial_t\widehat p-L_\eta^*\widehat p$, where
$L_\eta^*$ is the forward (Fokker--Planck) operator. Then,
locally in $\eta$,
\begin{equation}\label{eq:calibration_stability}
|\eta-\eta^\star|^2\le C\left(
 \|\widehat p_0-p_0\|_{L^2(\Omega)}^2
 +\|R^\eta\|_H^2
 +\|\widehat p-p^{\eta^\star}\|_H^2\right).
\end{equation}
\end{proposition}

The proposition separates two sources of error: approximation of the forward equation and identification of the parameters from the induced densities. The latter depends on the observation design and cannot be inferred from the PDE alone. We now consider the associated optimal control problem.

\begin{example}[Calibration on synthetic data]
Figure~\ref{fig:ddpinn} illustrates calibration on the seasonal model of Section~\ref{sec:numerics} with assumed historical observation of prices. The plot shows the learning dynamics and the reconstructed coefficients in the experiment.
\begin{figure}[htbp]
    \centering
    \includegraphics[width=0.90\linewidth]{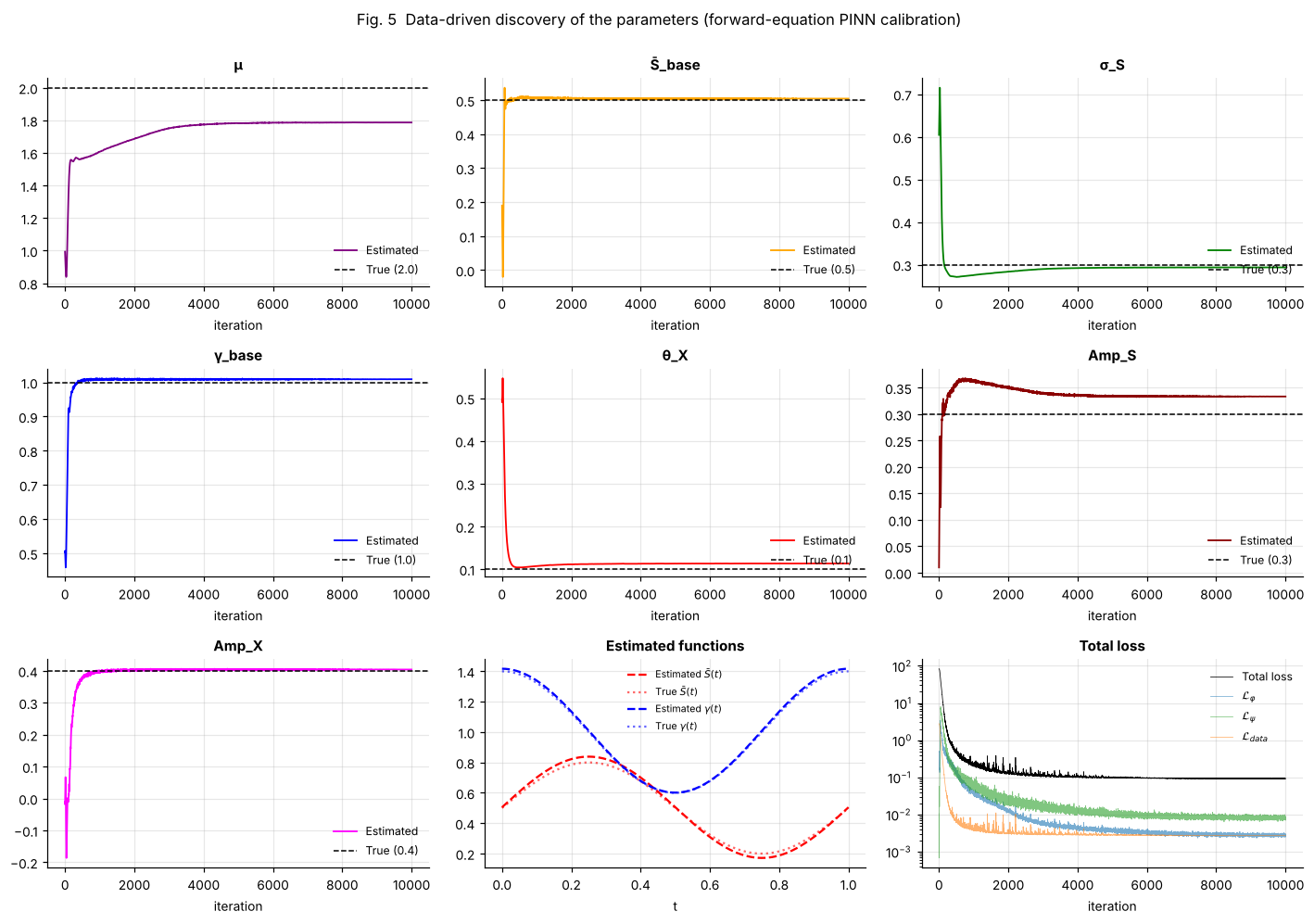}
    \caption{Parameter trajectories and reconstructed dynamics from the PINN calibration experiment}
    \label{fig:ddpinn}
\end{figure}
\end{example}

\section{Model-based control: HJB and policy iteration}
\label{sec:optimal_control}

We next consider a benchmark problem in which the current market price and the model coefficients are known. This complete-information formulation leads to an HJB equation and provides a reference policy against which an estimated-price strategy may be compared. Under incomplete information, the same numerical method can be used as a plug-in approximation, but it does not give the HJB equation of the original partially observed problem.

\subsection{Full-information dynamic programming}

Let $\mathcal A^{\rm full}$ denote the class of admissible policies adapted to the filtration generated by the full state $(S,X,Y)$. For this benchmark, define the value function
\begin{equation}
\label{eq:value_function}
\begin{split}
V(t,s,x,y)=\sup_{\alpha\in\mathcal A^{\rm full}}
\mathbb E_{t,s,x,y}^{\alpha}\Bigg[
\int_t^T \Big(\beta S_\tau
\mathbf1_{\{\alpha_\tau\ge S_\tau\}}
+f(\alpha_\tau,S_\tau,X_\tau)\Big)\,d\tau
-g(Y_T)\Bigg],
\end{split}
\end{equation}
where $\mathbb E_{t,s,x,y}^{\alpha}$ denotes expectation conditional on $(S_t,X_t,Y_t)=(s,x,y)$ under the policy $\alpha$. This value represents the maximal expected reward over $[t,T]$ for a given initial state.

Under the usual well-posedness and dynamic programming assumptions, the value function satisfies the following principle.


\medskip
\noindent\textbf{Dynamic programming principle.} For any stopping time $\tau$ valued in $[t,T]$ with respect to the full-information filtration,
\[
V(t,s,x,y)
=
\sup_{\alpha\in\mathcal A^{\rm full}}
\mathbb E
\left[
\int_t^\tau
r(u,S_u,X_u,Y_u,\alpha_u)\,du
+
V(\tau,S_\tau,X_\tau,Y_\tau)
\right],
\]
where
\[
r(t,s,x,y,\alpha)
=
\beta s
\mathbf1_{\{\alpha\ge s\}}
+
f(\alpha,s,x).
\]


This is the standard dynamic programming identity for controlled diffusions; see \citep{FlemingSoner2006} and \citep{pham2009continuous}.

\subsection{The Hamilton--Jacobi--Bellman equation}

As in the linear--quadratic benchmark, we define the Hamiltonian using generic arguments $p=(p_s,p_x,p_y)\in\mathbb R^3$ and $M\in\mathbb S^3$:
\begin{equation}\label{eq:hamiltonian_general}
\begin{aligned}
&H(t,s,x,y,p,M)
=\mu(t,s)p_s+\frac12\sigma^2(t,s)M_{ss}\\
&\quad+\sup_{a\in A}\Big\{
 f(a,s,x)+\beta s\,\mathbf1_{\{a\ge s\}}
 +\gamma(t,x,s,a)p_x\\
&\hspace{38mm}+\frac12\theta^2(t,x,s,a)M_{xx}
 -(1+\beta)s\,\mathbf1_{\{a\ge s\}}p_y
\Big\}.
\end{aligned}
\end{equation}
Since the Brownian motions driving $S$ and $X$ are independent, no mixed second-order derivatives appear in \eqref{eq:hamiltonian_general}. Under the usual dynamic programming assumptions, the corresponding HJB equation is
\begin{equation}\label{eq:HJB_general}
\left\{
\begin{aligned}
&V_t+H(t,s,x,y,DV,D^2V)=0,\\
&V(T,s,x,y)=-g(y).
\end{aligned}
\right.
\end{equation}
The maximization in \eqref{eq:hamiltonian_general} accounts for the campaign reward, the margin earned on successful transactions, and the effect of expenditure on future value. The gradient and Hessian in \eqref{eq:HJB_general} are evaluated at $(s,x,y)$.

If $V$ is smooth, a verification theorem applies, and the pointwise maximum is attained, an optimal feedback control may be chosen from
\begin{equation}\label{eq:optimal_feedback_general}
\begin{aligned}
\alpha^*(t,s,x,y)\in\underset{a\in A}{\operatorname{argmax}}\;\Big\{
& f(a,s,x)+\beta s\,\mathbf1_{\{a\ge s\}}\\
&+\gamma(t,x,s,a)V_x
+\frac12\theta^2(t,x,s,a)V_{xx}\\
&-(1+\beta)s\,\mathbf1_{\{a\ge s\}}V_y
\Big\}.
\end{aligned}
\end{equation}
The terms involving $V_s$ and $V_{ss}$ belong to the full Hamiltonian but do not affect the pointwise maximization because the price dynamics are uncontrolled.

Under incomplete information, however, the relevant state includes the conditional distribution $\mathcal L(S_t\mid\mathcal G_t^\alpha)$ of the hidden price. Because the observation mechanism depends on the policy, this conditional distribution has controlled filtering dynamics. The separated problem is consequently measure-valued. Replacing $S_t$ by $\widehat S_t^N$ in the benchmark equation is an approximation rather than an exact separation principle.

The value function need not be smooth, and the viscosity-solution framework is natural for an HJB equation of this type; see \citep{crandall1992user}. The discontinuity of the winning indicator, together with possible loss of attainment at a threshold, requires care in establishing a rigorous comparison or verification principle. We do not address this technical question here, and use \eqref{eq:HJB_general} as the basis for numerical approximation.

\subsection{Policy iteration with physics-informed neural networks}

To translate the control formulation into a bidding rule, one needs to approximate the value function and perform the pointwise maximization in the HJB equation. Grid-based schemes may become expensive for several state variables. We therefore consider a physics-informed neural network (PINN) approximation of the value function.

Let
\[
V_\theta(t,s,x,y)
\]
be a neural-network approximation of $V$, with trainable parameters $\theta$.

The terminal condition is enforced through the parameterization
\[
V_\theta(t,s,x,y)
=
(T-t)\,
N_\theta(t,s,x,y)
-
g(y),
\]
where $N_\theta$ is an unconstrained neural network. A strict budget constraint $Y_t\ge0$, if imposed, would also require an appropriate boundary condition at $y=0$.

We use a generalized policy-iteration procedure, alternating between
\begin{enumerate}
    \item policy evaluation, during which the value function is updated for a fixed policy;
    \item policy improvement, during which the policy is updated using the current approximation of the value function.
\end{enumerate}

At each policy-evaluation step, automatic differentiation provides the first- and second-order derivatives entering the linear equation. The policy-improvement step maximizes the Hamiltonian computed from the current network.

Algorithm~\ref{alg:pi_pinn} summarizes the computational steps.

\begin{algorithm}[H]
\caption{Policy iteration with a PINN value-function approximation}
\label{alg:pi_pinn}

\begin{algorithmic}[1]

\STATE Initialize value network $V_\theta$

\REPEAT

\STATE Policy evaluation

\STATE Fix the current policy and solve the corresponding linear policy-evaluation equation approximately with $V_\theta$

\STATE Sample collocation points

\STATE Evaluate the HJB residual using automatic differentiation

\STATE Update $\theta$ by stochastic gradient descent

\STATE Policy improvement: maximize the Hamiltonian pointwise using the current derivatives of $V_\theta$

\UNTIL convergence

\end{algorithmic}

\end{algorithm}

The procedure follows the policy-iteration principle, but viscosity-solution theory alone does not establish convergence of the neural approximation. In particular, approximation, optimization, and policy-improvement errors must be analysed separately. We assess the numerical behavior of the method below.

\section{Policy learning through reinforcement learning}
\label{sec:reinforcement_learning}

The HJB approach relies on a specified and calibrated dynamics model. A practical alternative is to update bidding policies from observed actions and rewards, without first estimating every model coefficient. We describe this second approach using randomized controls and an entropy-regularized actor--critic approximation. Both methods must respect the same information constraint if their outcomes are to be compared.

\subsection{Randomized policies and relaxed controls}

Instead of selecting a single bid, the controller chooses at time $t$ a probability measure
\[
\pi_t\in\mathcal P(A),
\]
and the bid is sampled according to
\[
\alpha_t\sim \pi_t.
\]
The process $\pi=(\pi_t)_{t\in[0,T]}$ is progressively measurable and takes values in $\mathcal P(A)$. The continuous-time model below uses a relaxed (measure-valued) control; averaging the infinitesimal drift and variance should not be confused with replacing the diffusion coefficient by its ordinary average.

For a probability measure $\pi\in\mathcal P(A)$, define
\[
\widetilde\gamma(t,x,s,\pi)
=
\int_A
\gamma(t,x,s,a)\pi(da),
\]
and
\[
\widetilde\theta(t,x,s,\pi)
=
\left(
\int_A
\theta^2(t,x,s,a)\pi(da)
\right)^{1/2}.
\]
The relaxed state dynamics are then written as
\[
\left\{
\begin{aligned}
dS_t^\pi
&=
\mu(t,S_t^\pi)dt
+
\sigma(t,S_t^\pi)dW_t^1,
\\
dX_t^\pi
&=
\widetilde\gamma(t,X_t^\pi,S_t^\pi,\pi_t)dt
+
\widetilde\theta(t,X_t^\pi,S_t^\pi,\pi_t)dW_t^2.
\end{aligned}
\right.
\]

The expected budget expenditure under the relaxed policy is
\[
dY_t^\pi
=
-(1+\beta)S_t^\pi
\int_A
\mathbf1_{\{a\ge S_t^\pi\}}
\pi_t(da)\,dt.
\]

Similarly, the average trading revenue is
\[
R(t,S_t^\pi,\pi_t)
=
\beta S_t^\pi
\int_A
\mathbf1_{\{a\ge S_t^\pi\}}
\pi_t(da).
\]

This formulation is related to entropy-regularized stochastic control and continuous-time reinforcement learning; see, for example, \citep{JMLR:v21:19-144}, \citep{JMLR:v23:21-0947}, and \citep{jia_q-learning_2022}.

\subsection{Entropy-regularized objective}

To favor exploration, consider the entropy penalty associated with a policy density $\pi_t(a)$ on $A$:
\[
\mathcal E(\pi_t)
=
\int_A
\pi_t(a)\log\pi_t(a)\,da.
\]

For an entropy temperature $\tau_{\rm ent}>0$ (distinct from the exploration rate $\lambda$), the criterion becomes
\begin{equation}
\label{eq:rl_objective}
\sup_{\pi}
\mathbb E
\left[
\int_0^T
\left(
f_\pi(t,X_t^\pi,S_t^\pi)
+
R(t,S_t^\pi,\pi_t)
-
\tau_{\rm ent}\mathcal E(\pi_t)
\right)dt
-
g(Y_T^\pi)
\right],
\end{equation}
where
\[
f_\pi(t,x,s)
=
\int_A
f(a,s,x)\pi_t(da).
\]

The parameter $\tau_{\rm ent}$ determines the relative weight assigned to entropy. Larger values favor more diffuse bidding distributions. Because entropy is differential entropy, its numerical value also depends on the units of the bid; the action range is fixed here.

\subsection{Actor--critic approximation}

For numerical implementation, we use an actor--critic scheme on a time grid $0=t_0<\cdots<t_K=T$. The information fed to the networks must be observable. Let $O_k$ be a feature vector that may contain the estimated price $\widehat S_{t_k}^N$, observed performance and budget, or a finite-dimensional approximation of the conditional price distribution. The actor $\pi_\psi(\cdot\mid O_k)$ represents a policy density on $A$ (for example, a truncated Gaussian), while the critic $J_\theta(t_k,O_k)$ approximates its value.

For an action $a_k$ chosen at time $t_k$, let $r_k$ be the observed reward earned over $[t_k,t_{k+1}]$, including the entropy regularisation. The terminal condition is $J_\theta(T,O_K)=-g(Y_T)$, with no additional terminal term in $r_{K-1}$. Define the temporal-difference innovation
\[
\delta_k=r_k+J_\theta(t_{k+1},O_{k+1})
 -J_\theta(t_k,O_k),
\]
For an exact continuation value and a sufficient observation state, the temporal-difference increment has zero conditional mean. In a finite-dimensional feature approximation one instead seeks approximately
\[
\mathbb E[\delta_k\mid O_k]\simeq0.
\]
One possible critic loss is therefore the squared conditional expectation
\[
\mathcal L_{\rm critic}(\theta)=
\sum_{k=0}^{K-1}
\mathbb E\big[\big|\mathbb E[\delta_k\mid O_k]\big|^2\big].
\]
With an advantage estimator $\widehat A_k$, the actor update takes the form
\[
\psi\leftarrow\psi+\eta_\psi\sum_{k=0}^{K-1}
\widehat A_k\nabla_\psi\log\pi_\psi(a_k\mid O_k).
\]
This condition does not require the realised temporal-difference error to vanish pathwise. The above discretisation defines an implementable scheme, but its convergence to a continuous-time optimal policy is not established here.

Unless $O_k$ is a sufficient statistic for the observation history, this reduction also introduces a state-approximation error. Algorithm~\ref{alg:ct_rl} summarizes the implementation.

\begin{algorithm}[H]
\caption{Discrete actor--critic approximation of a continuous-time control problem}
\label{alg:ct_rl}
\begin{algorithmic}[1]
\STATE Initialize actor parameters $\psi$ and critic parameters $\theta$
\REPEAT
    \STATE Simulate trajectories under the randomized policy $\pi_\psi$
    \STATE Compute rewards and budget updates
    \STATE Update the critic by minimizing the temporal-difference residual
    \STATE Update the actor using the entropy-regularized policy gradient
\UNTIL convergence
\RETURN Learned randomized policy $\pi_\psi$
\end{algorithmic}
\end{algorithm}

Unlike the HJB-based strategy, this approach does not require an explicit parameterization of the drift and volatility coefficients. It does, however, require sufficiently informative observations and rewards. We compare its  numerical behavior with that of the model-based approximation in Section~\ref{sec:numerics}.

\section{Numerical experiments and operational interpretation}
\label{sec:numerics}

\subsection{Experimental design}

We consider the following seasonal diffusion model, together with a performance criterion consistent with the payment rule of Section~\ref{sec:general_model}:
\begin{equation}\label{dynamics}
\begin{aligned}
dS_t&=\mu(\bar S(t)-S_t)\,dt+\sigma_S\,dW_t^S,\\
\bar S(t)&=\bar S_{\rm base}+\mathrm{Amp}_S\sin(2\pi ft),\\
dX_t&=\gamma(t)\alpha_t\mathbf1_{\{\alpha_t\ge S_t\}}\,dt
 +\theta_X\,dW_t^X,\\
\gamma(t)&=\gamma_{\rm base}(1+\mathrm{Amp}_X\cos(2\pi ft)),\\
dY_t&=-(1+\beta)S_t\mathbf1_{\{\alpha_t\ge S_t\}}\,dt,\\
J(\alpha)&=\mathbb E\bigg[\int_0^T
\Big(X_t+\beta S_t\mathbf1_{\{\alpha_t\ge S_t\}}\\
&\hspace{3cm}-\kappa(1+\beta)^2\alpha_t^2
\mathbf1_{\{\alpha_t\ge S_t\}}\Big)dt-Y_T^2\bigg].
\end{aligned}
\end{equation}
The quadratic bidding penalty is included in the running reward. Since the Ornstein--Uhlenbeck price is not almost surely positive, a positivity-preserving specification or a documented truncation would be needed for a literal auction-price interpretation.
The parameters used in the  experiments are collected in Table~\ref{tab:parameters}.

\begin{table}[htbp]
    \centering
    \begin{tabular}{cc}
        \toprule
        Parameter & Value \\
        \midrule
        $\mu$ & $2.0$ \\
        $\overline{S}_{\mathrm{base}}$ & $0.5$\\
        $\mathrm{Amp}_S$ & $0.3$\\
        $\sigma_S$ & $0.3$ \\
        $\gamma_{\mathrm{base}}$ & $1.0$ \\
        $\mathrm{Amp}_X$ & $0.4$\\
        $\theta_X$ & $0.1$ \\
        $f$ (seasonal frequency) & $1.0$ \\
        $\kappa$ & $0.5$\\
        $\beta$ & $0.2$ \\
        $B$ & $1.0$ \\
        $\lambda$ & $3.0$ \\
        $d$ & $0.1$ \\
        \bottomrule
    \end{tabular}
    \caption{Parameters used in the seasonal-price simulations. $d$ is the share of budget allocated to price estimation.}
    \label{tab:parameters}
\end{table}

\subsection{Model-based control with observed and estimated prices}

\begin{figure}[htbp]
    \centering
    \includegraphics[width=0.86\linewidth]{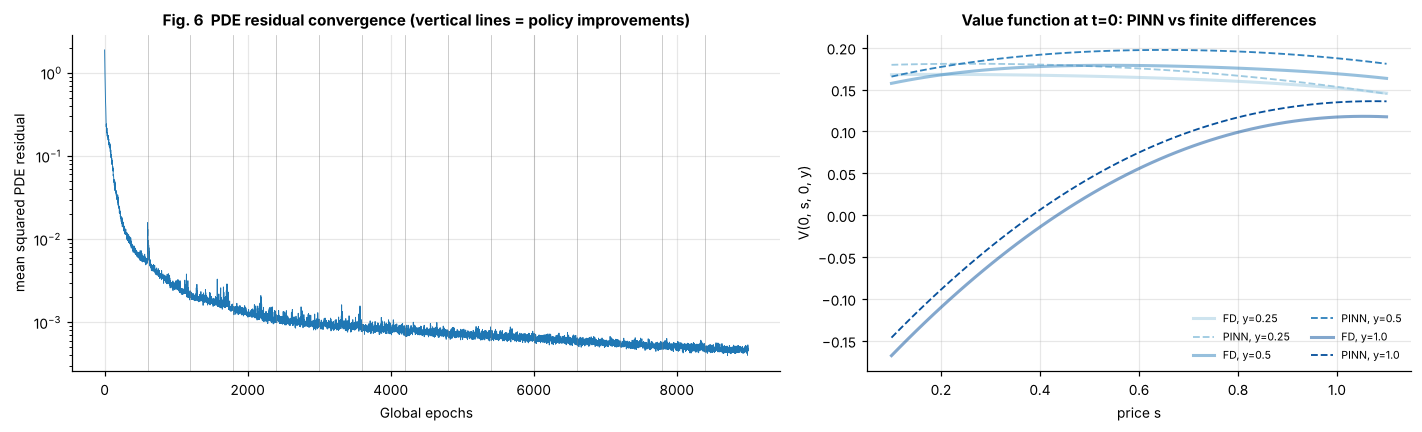}
    \caption{PDE residual during the PINN policy-iteration computation}
    \label{fig:loss_GPI}
\end{figure}

Figure~\ref{fig:loss_GPI} displays the evolution of the PDE residual during the  policy-iteration procedure.

\begin{figure}[htbp]
    \centering
    \includegraphics[width=0.90\linewidth]{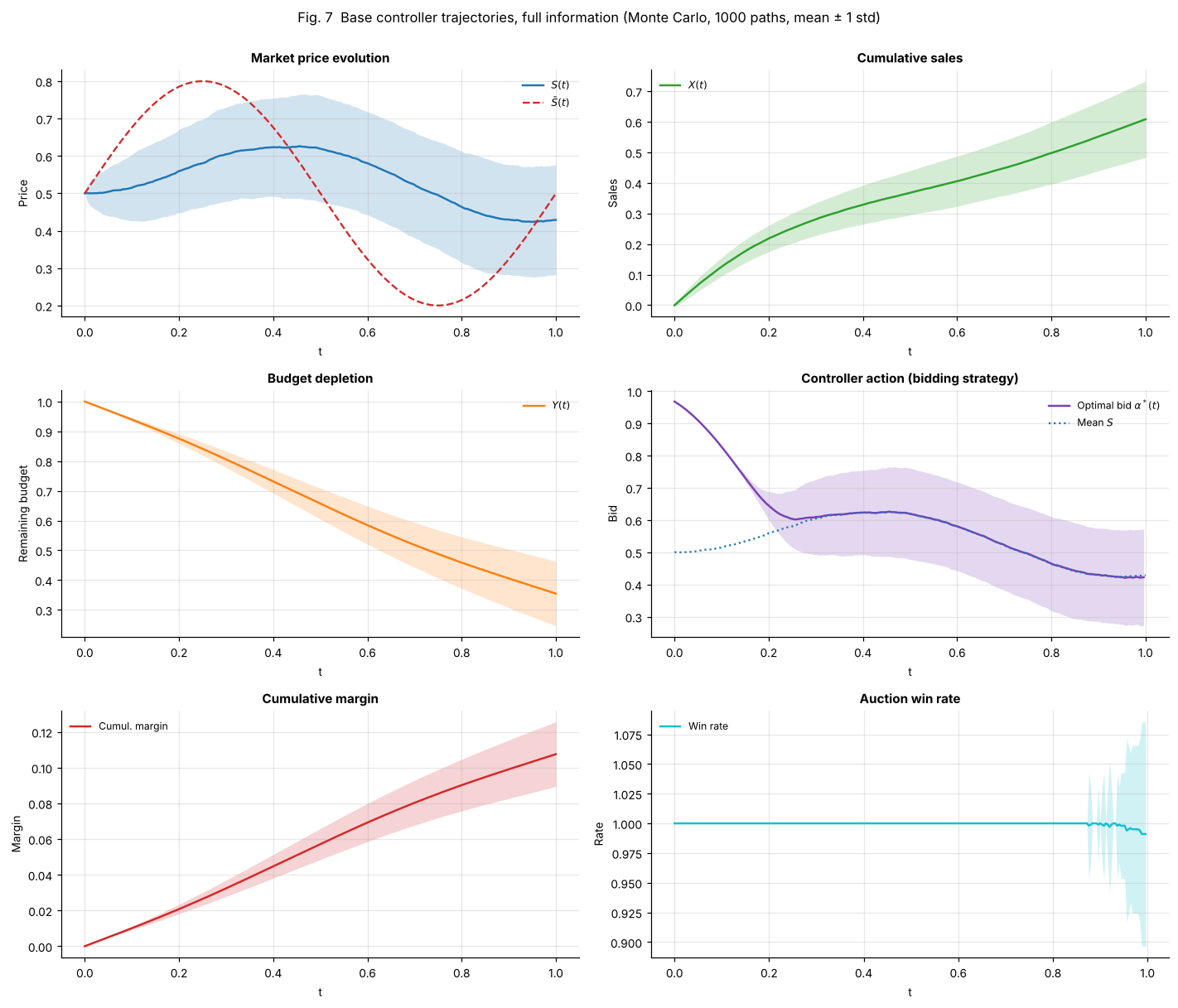}
    \caption{State and bidding trajectories from the full-information PINN implementation}
    \label{fig:traj}
\end{figure}

The full-information benchmark yields the trajectories in Figure~\ref{fig:traj}, illustrating the interaction between the market price, bids, performance, and remaining budget.

\begin{figure}[htbp]
    \centering
    \includegraphics[width=0.90\linewidth]{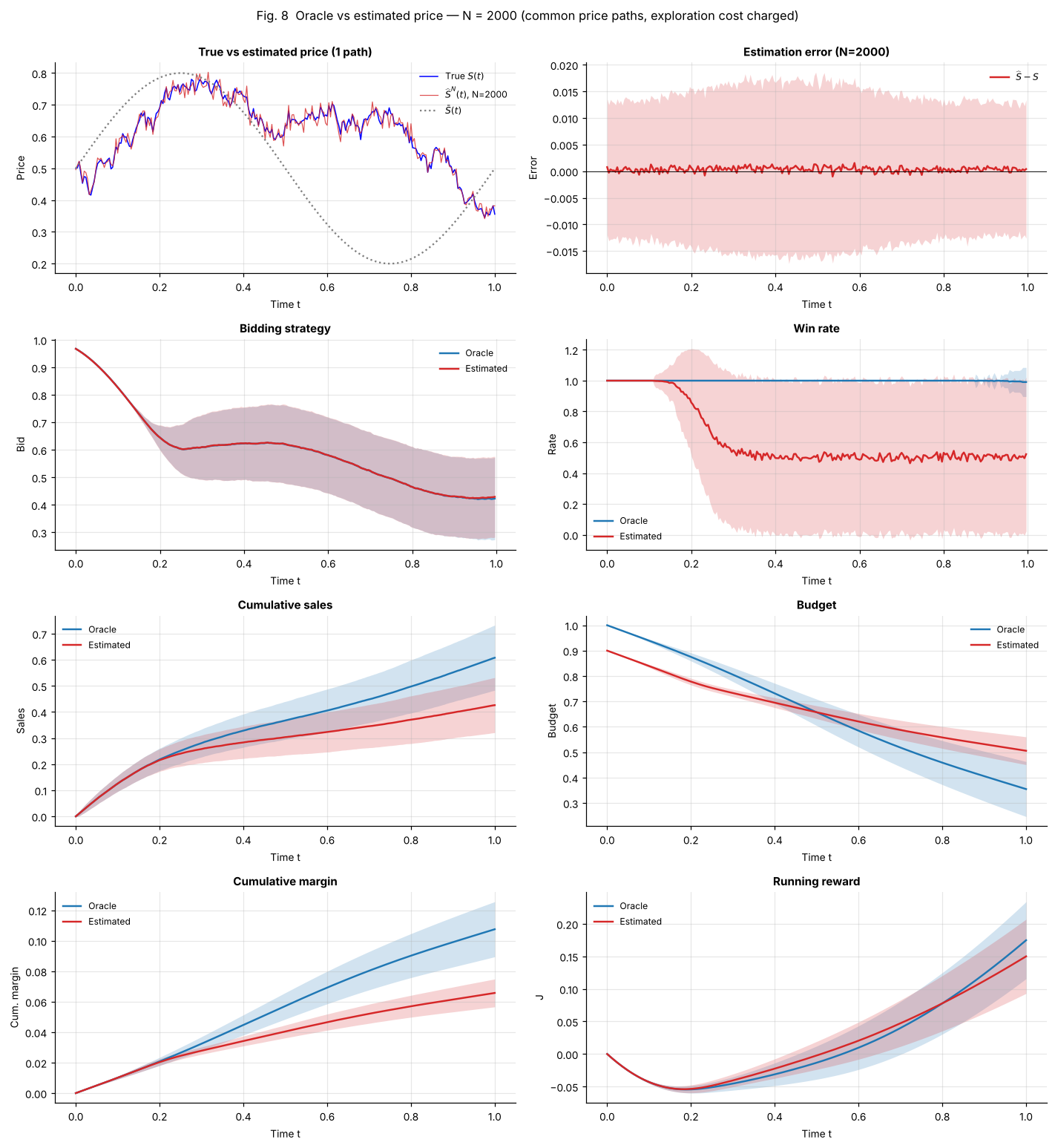}
    \caption{Estimated-price trajectories: prices, bidding, win frequency, performance, trading margin, and budget}
    \label{fig:traj_est}
\end{figure}

Figure~\ref{fig:traj_est} compares the trajectories with and without price estimation, using an exploratory allocation corresponding to $10\%$ of the initial budget. 

\subsection{Randomized actor--critic policy}

\begin{figure}[htbp]
    \centering
    \includegraphics[width=0.90\linewidth]{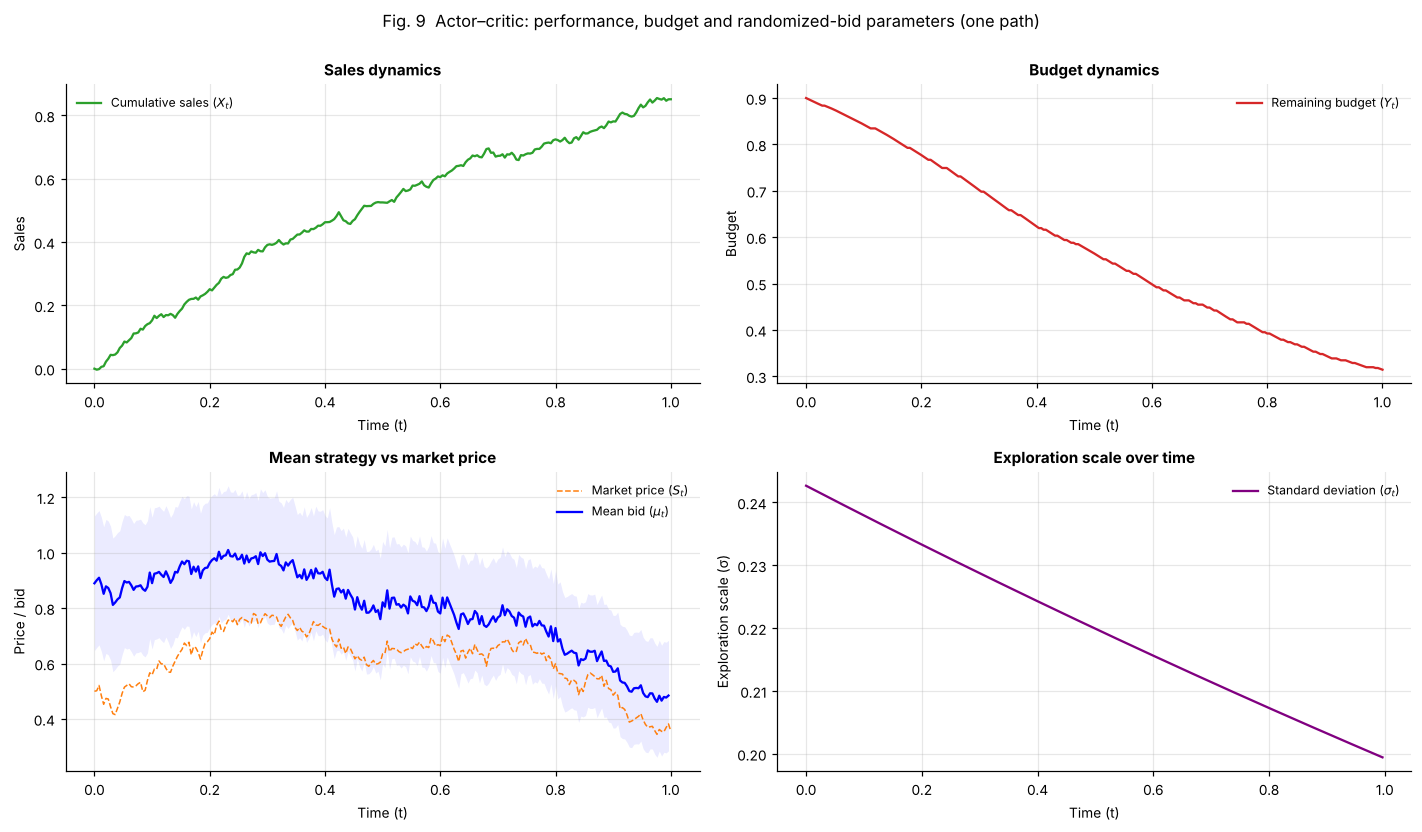}
    \caption{Performance, budget expenditure, and randomized-bid parameters from the actor--critic experiment}
    \label{fig:stats_RL}
\end{figure}

Figure~\ref{fig:stats_RL} illustrates the evolution of the randomized-policy parameters, together with performance and expenditure under reinforcement learning.

\subsection{A budget-pacing benchmark}
\label{subsec:pid}

To compare the model-based and learning-based policies with a rule closer to
operational practice, we consider a budget-pacing heuristic based on a
proportional--integral--derivative (PID) controller. The rule does not attempt
to solve the optimization problem \eqref{eq:control_problem}. Its only objective is to keep the
expenditure of the campaign close to a prescribed spending schedule.

Let $Y^{\mathrm{ref}} = (Y^{\mathrm{ref}}_t)_{t\in[0,T]}$ be a deterministic
target for the remaining budget, decreasing from $B_0$ at time $0$. In the
 experiments we use uniform pacing,
\begin{equation}
  Y^{\mathrm{ref}}_t = B_0\Bigl(1 - \frac{t}{T}\Bigr),
  \label{eq:pid-ref}
\end{equation}
although a seasonal schedule following the expected price level could be used
instead. The tracking error is
\begin{equation}
  e_t = Y_t - Y^{\mathrm{ref}}_t .
  \label{eq:pid-error}
\end{equation}
A positive error indicates underspending relative to the schedule, and a
negative error indicates overspending.
Given a baseline bid $\bar\alpha \in A$ and nonnegative gains
$K_P$, $K_I$, $K_D$, the PID bid is
\begin{equation}
  \alpha^{\mathrm{PID}}_t
  = \operatorname{proj}_{[0,A_{\max}]}
    \Bigl( \bar\alpha + K_P\, e_t + K_I \int_0^t e_u\, du + K_D\, \dot e_t \Bigr).
  \label{eq:pid-bid}
\end{equation}
Underspending therefore raises the bid, which increases the probability of
winning auctions, and overspending lowers it. The integral term corrects
persistent deviations from the schedule. The derivative term damps the
response to rapid changes in expenditure.

Since $Y$ is absolutely continuous, $\dot e_t$ exists for almost every $t$,
but it inherits the discontinuities of the winning indicator. In practice, the
rule is implemented on the time grid $0 = t_0 < \dots < t_K = T$ of
Section~6.3:
\begin{equation}
  \alpha^{\mathrm{PID}}_{t_k}
  = \operatorname{proj}_{[0,A_{\max}]}
    \Bigl( \bar\alpha + K_P\, e_k + K_I \sum_{j\le k} e_j\, \Delta t
      + K_D\, \frac{e_k - e_{k-1}}{\Delta t} \Bigr),
  \label{eq:pid-discrete}
\end{equation}
where $e_k = e_{t_k}$ and $\Delta t$ is the grid step. When the bid saturates
at $0$ or $A_{\max}$, the accumulated error may grow without improving the
tracking. A standard anti-windup correction, freezing the integral term during
saturation, prevents this effect.

The discrete-time PID rule \eqref{eq:pid-discrete} depends only on the current
and past values of the observed remaining budget. It can
therefore be implemented without access to the latent market
price and requires neither the price estimator $\widehat S^N$
nor expenditure on exploratory subcampaigns.

Budget pacing reflects a common practice in media trading: ensuring that the
client's budget is spent regularly over the campaign horizon. The rule is
transparent, easy to tune, and robust to misspecification of the price
dynamics. Its limitations follow from the same simplicity. It ignores the
market price, campaign performance and the agency margin, and it does not
account for the quadratic bidding cost in the criterion. In particular, it may
win many auctions at unfavorable prices, which increases performance but not
necessarily the value of \eqref{eq:criterion}. The rule is therefore a useful operational
benchmark rather than a competing solution of the control problem. The gains
$(K_P, K_I, K_D)$ and the baseline bid $\bar\alpha$ used in the
experiments are given in Table~\ref{tab:pid}.

\begin{table}[ht]

  \centering
  \begin{tabular}{ll}
    \toprule
    Parameter & Value \\
    \midrule
    $\overline{\alpha}$ & $0.5$  \\
    $K_p$ & $2.0$  \\
    $K_i$ & $4.0$  \\
    $K_d$ & $0.05$ \\
    Time step & $\Delta t = T/250$  \\
    Bounds & $[0, A_{\max}] = [0, 1.5]$  \\
    \bottomrule
  \end{tabular}
  \caption{Parameters of the PID budget-pacing rule}
  \label{tab:pid}
\end{table}

\subsection{Outcomes and comparison}

\begin{figure}[htbp]
    \centering
    \includegraphics[width=0.9\linewidth]{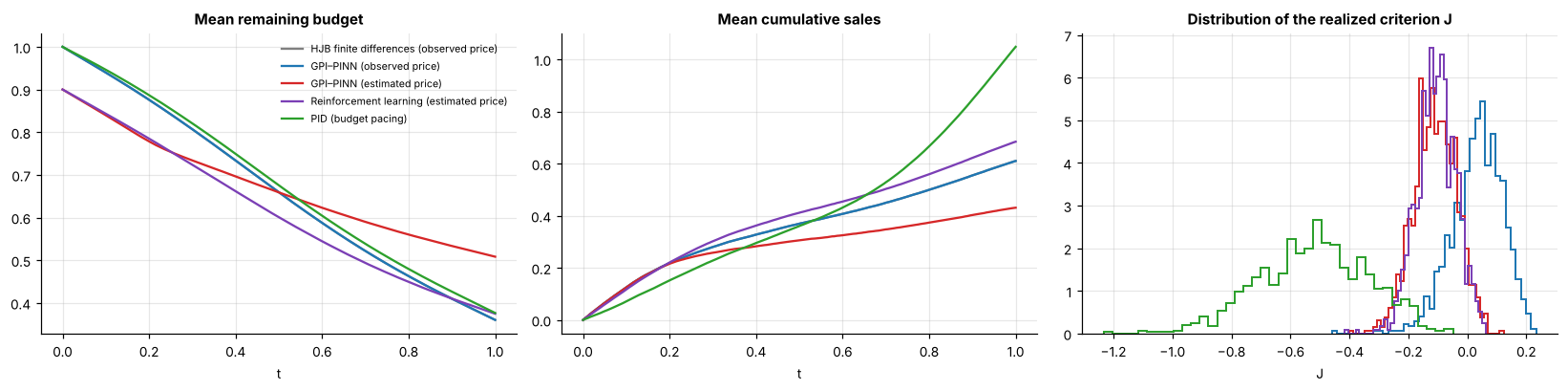}
    \caption{Performance comparison between the different methods on 1000 common trajectories}
    \label{fig:comparison}
\end{figure}

Figure~\ref{fig:comparison} gathers the outcomes of five bidding rules:
a finite-difference approximation of the HJB benchmark and the GPI--PINN policy,
both with observed prices; the GPI--PINN feedback evaluated at the estimated
price $\widehat S^N$; the actor--critic policy of Section~6, fed with the same
estimated price; and a PID budget-pacing rule. The PID rule adjusts the bid so
that cumulative expenditure tracks a prescribed spending schedule. It uses
neither the value function nor the price estimator, and serves as an
operational heuristic rather than as a solution of ~\eqref{eq:control_problem}.

The three panels display the mean remaining budget, the mean cumulative
performance, and the empirical distribution of the realized criterion. Several
features are visible. First, the policies based on the estimated price start
from a budget of $0.9$, since $10\%$ of the initial budget is allocated to the
exploratory subcampaigns. Second, the observed-price policies spend their
budget at a comparable rate and end the horizon with approximately one third of
the initial budget. The estimated-price GPI--PINN policy spends more slowly and
ends with roughly half of it. Third, the PID rule produces the largest
cumulative performance, whereas the estimated-price GPI--PINN policy produces
the smallest.

The distribution of the realized criterion reverses this ordering. The GPI--PINN value with observed prices exceeds
the other  values, whereas the PID rule yields more sales together with
a markedly lower and more dispersed criterion. This is consistent with the
structure of the objective: the PID rule tracks expenditure without accounting
for the quadratic bidding cost or for the margin earned on each transaction.
The policies based on the price signal lie between these two cases.

The gap between the observed-price and estimated-price GPI--PINN policies has
two distinct sources. The first is the exploration cost, which reduces the
budget available to the main campaign. The second is the threshold effect
discussed after Proposition~4.2. A feedback calibrated to bid close to the
estimated price wins with high probability in the proxy model, but only with
probability close to one half in the true model when
$\widehat S^N_t - S_t$ is approximately centred. Figure~8 shows this
reduction in the win rate. It explains both the slower expenditure and the
lower performance of the estimated-price policy. Consequently, the observed
gap measures the cost of using a plug-in policy, not only the cost of
estimating the price.

\section{Discussion and conclusion}
\label{sec:conclusion}

We have formulated media trading as a dynamic bidding and budget-allocation problem in which the agency sees market-clearing prices only after successful auctions. The model makes the three components of the agency's decision explicit: campaign performance, trading margin and the use of the client's remaining budget. An additional feature is the endogenous information structure, as the bids determine which clearing prices are observed.

Randomized exploratory subcampaigns provide a simple statistical instrument for recovering a price signal. Under the specified cross-sectional model, the estimation error admits a mean-square bound of order $N^{-1}$. We then consider two ways to turn price and performance information into bidding policies: a calibrated stochastic model with an HJB/PINN approximation, and an entropy-regularized actor--critic approach. These methods offer complementary computational routes rather than new results in stochastic control theory.

\paragraph{Operational implications.}
The formulation highlights why good bidding decisions cannot be judged by conversions alone. A policy generating more sales may also erode the agency's margin or exhaust the client's budget too early. Likewise, exploration has a potential value through improved price information but entails a cost in advertising expenditure. These trade-offs motivate the use of a common, transparent objective when evaluating competing policies.

\paragraph{Limitations and outlook.}
Our estimator result relies on an independent exploratory sampling scheme and does not establish optimality for control with censored data. The HJB equation is a full-information benchmark, and substituting an estimated price leads to a plug-in policy. The exploratory budget is not jointly optimized with the main campaign. 

\appendix

\section{Expression of \texorpdfstring{$Z^\pi$}{Zpi} as a diffusion process}
\label{app:zpi-diffusion}

Let $A=[0,A_{\max}]$ and suppose $S_t\in(0,A_{\max})$.
Let $\Pi_t(da)=\pi_t(a)da$ be a possibly random family of
probability measures with $\pi_t(a)>0$ on $(0,A_{\max})$.
Suppose $\pi_t(a)$ is $C^1$ in $a$ and absolutely continuous
with finite variation in time,
\[
\pi_t(a)=\pi_0(a)+\int_0^t\dot\pi_u(a)\,du,
\qquad \int_A\dot\pi_t(a)\,da=0,
\]
with sufficient integrability to apply the It\^o formula for a
random finite-variation field. Define the randomized-policy winning
probability
\[
Z_t^\pi=\int_{S_t}^{A_{\max}}\pi_t(a)\,da.
\]
Since $\partial_s Z=-\pi_t(s)$ and
$\partial_{ss}Z=-\partial_s\pi_t(s)$, It\^o's formula yields
\begin{equation}\label{eq:Zpi_corrected}
\begin{split}
dZ_t^\pi={}&\left[\int_{S_t}^{A_{\max}}\dot\pi_t(a)\,da
-\mu(t,S_t)\pi_t(S_t)
-\frac12\sigma^2(t,S_t)\partial_s\pi_t(S_t)\right]dt\\
&-\sigma(t,S_t)\pi_t(S_t)\,dW_t^1.
\end{split}
\end{equation}
The second-order term therefore has a negative sign.

Because $\pi_t$ is strictly positive on the price interval,
$s\mapsto\Pi_t([s,A_{\max}])$ is strictly decreasing on that
interval. For $z\in(0,1)$ let $s(\Pi_t,z)$ denote its inverse.
Then $S_t=s(\Pi_t,Z_t^\pi)$ and \eqref{eq:Zpi_corrected}
can be rewritten entirely in terms of $(\Pi_t,Z_t^\pi)$.
Atomlessness alone is not sufficient for invertibility (a
non-atomic probability density may vanish on an interval).
Outside the support of $\Pi_t$, $Z_t^\pi$ equals zero or one,
so the latent price cannot be recovered by inversion.
The finite-variation condition is important here: if the policy distribution had a martingale component, the formula would include additional It\^o--Wentzell terms.

\section{Proof of the Calibration Consistency Bound}
\label{app:loss-consistency}

We prove Proposition~\ref{prop:calibration_consistency}. Denote by $F(\eta)=p^\eta$ the parameter-to-density map, with values in $H=L^2((0,T)\times\Omega)$. By Fr\'echet differentiability,
\[
F(\eta)-F(\eta^\star)
=DF(\eta^\star)(\eta-\eta^\star)+
 o(|\eta-\eta^\star|).
\]
The assumed injective derivative with a quantitative lower
bound implies, for $\eta$ in a sufficiently small neighborhood,
\begin{equation}\label{eq:local_identifiability}
|\eta-\eta^\star|\le\frac{2}{c_0}
\|p^\eta-p^{\eta^\star}\|_H.
\end{equation}
The estimate is local and relies on the quantitative injectivity assumption.

Let $U_\eta(t,s)$ denote the forward evolution family of
$\partial_t p=L_\eta^*p$, assumed uniformly bounded by $M$
on $0\le s\le t\le T$. Duhamel's formula for the approximate
density $\widehat p$ gives
\[
\widehat p(t)-p^\eta(t)
=U_\eta(t,0)(\widehat p_0-p_0)
+\int_0^tU_\eta(t,s)R^\eta(s)\,ds,
\qquad R^\eta=\partial_t\widehat p-L_\eta^*\widehat p.
\]
By Cauchy--Schwarz and the uniform bound on $U_\eta$,
\[
\|\widehat p-p^\eta\|_H^2\le C_T
\left(\|\widehat p_0-p_0\|_{L^2(\Omega)}^2
+\|R^\eta\|_H^2\right).
\]
Finally,
\[
\|p^\eta-p^{\eta^\star}\|_H
\le\|p^\eta-\widehat p\|_H
+\|\widehat p-p^{\eta^\star}\|_H,
\]
and the claimed estimate follows. For controlled dynamics, the forward operator must incorporate the experimental policy or the corresponding state--action occupation measure. If the parameter-to-density map is not locally identifiable, the same argument does not yield parameter consistency.

\section{Convergence of the Randomized Observation Process}
\label{app:approx}

Fix $t\in[0,T]$ and condition on $S_t=s$. By the conditional independence of exploratory bids and price noises, the variables $I_t^i=\mathbf 1_{\{\alpha^i\ge s+\varepsilon_t^i\}}$ are independent Bernoulli variables with success probability $h_t(s)$ given by \eqref{eq:observation_map}. Consequently,
\[
\mathbb E\big[|z_t^N-h_t(S_t)|^2\mid S_t\big]
=\frac{h_t(S_t)(1-h_t(S_t))}{N}\le\frac{1}{4N}.
\]
Taking expectations gives the fixed-time bound. Integration and Tonelli's theorem yield the integrated result without any temporal independence assumption.

To establish strict monotonicity, differentiate the Gaussian distribution function under the bounded-action integral in \eqref{eq:observation_map}. This gives \eqref{eq:observation_derivative}. On a compact price interval $K$, the derivative is bounded away from zero, and the clipped inverse in \eqref{eq:price-estimator} is Lipschitz. The price-estimation bound then follows as in Lemma~\ref{lemma1}.

We do not claim convergence in a process topology. A continuous-time interpretation of the threshold-based winning indicators requires an explicit temporal sampling model.

\small

\bibliographystyle{plain}
\bibliography{sample701}

\end{document}